\documentclass[12pt,thmsa]{article}
\usepackage{sw20lart}

\input tcilatex
\begin{document}

\author{Amjad Hussain Shah Gilani \\
National Center for Physics\\
Quaid-i-Azam University\\
Islamabad 45320, Pakistan\\
Email: ahgilani@yahoo.com}
\title{Mass relations among family members of quarks and leptons}
\date{}
\maketitle

\begin{abstract}
The various mass relations among members of quark and lepton families are
given. Three mass relations for the charm, beauty, and top quark family
members are given and three mass relations for the electron, muon, and tau
lepton family members are presented.
\end{abstract}

Quantum Choromodynamics (QCD), the theory of strong interactions, is unique
among physical theories in its combination of logical closure and empirical
success. In the asymptotic Bjorken limit, the study of the properties of
strong interactions has proved to be one of the most creative ideas \cite
{hep-ph/0502113}. In Regge asymptotics, the number of partons increases
rapidly due to QCD bremsstrahlung, relatively little work has been done.
This limit of strong interactions, which we shall call the Regge limit, was
studied intensively in the 60's and indeed led eventually to string theory 
\cite{hep-ph/0502190}. The rapid growth in the number of partons due to QCD
bremsstrahlung in the leading logrithmic approximation is described by the
BFKL equation \cite{BFKLeq}. There is a sense in which QCD is better than it
has to be. It is not a complete theory of the world. Even within the
standard model quarks have additional interactions, beyound QCD, that become
significant -- indeed, problematic -- at short distances \cite
{physics/0212025}. In the words of Professor Frank Wilczek, ``I find QCD
disturbing that it takes vast computer resources, careful limiting
procedures, to simulate the mass and properties of a proton with decent
accuracy. And for real time dynamics, like scattering, the situation appears
pretty hopeless. Nature, of course, gets such results fast and effortlessly.
But how, if not through some kind of computation, or a process we can mimic
by computation ? \cite{physics/0212025}''.

QCD offers a rich set of phenomena, which are not yet really understood. At
a more fundamental level, three properties of QCD are most striking. Two are
connected to the fermionic content of the theory: the breaking of chiral
symmetry and the axial anomaly. The first phenomenon gives rise to the
proton mass of nearly 1 GeV, while the quarks making it up have masses only
of the order of MeV. This can be understood as due to spontaneous breaking
of the approximate chiral symmetry of the lightest quarks and is also well
known in other models. The axial anomaly is connected to a true quantum
anomaly, and e.g. gives rise to the anomalous large mass of the $\eta
^{\prime }$ meson. The third property is confinement, the absence of the
colored degrees of freedom from the physical spectrum. It is this property
which significantly shapes the low-energy reality of daily life while
simultaneously being least understood of all the genuine non-perturbative
effects of QCD. At the same time it is one of the properties which has been
measured with the highest precision available: Free quarks have a unique
experimental signature due to their fractional electric charge. The absence
of such objects in nature has been established at a precision of the order
of 1:10$^{30}$. The true non-perturbative nature of QCD is the reason for
the complications in the study of these phenomena. Perturbation theory
describes strong effects reasonably well only at energies larger than a few
GeV, the precise scale depending on the process. At smaller energies, which
ultimately govern hadron and nuclear physics, the interaction is so strong
that perturbation theory is not applicable. Extensive results are available
from model calculations, e.g. A non-linear theory of strong interactions 
\cite{THRSkyrme}, The interacting gluon model \cite{hep-ph/0412293}, etc.
Recently, a new model for strong interactions is presented which establised
the concept from set theory and constrained from group theory \cite
{hep-ph/0404026,hep-ph/0410207,hep-ph/0501103,hep-ph/0502055,hep-ph/0502117}%
. In the color charge structure of gluons, a group property violation was
observed \cite{hep-ph/0404026} which become the cause to set up the new
model. This exactly predict the octet gluon structure and their charge
structure, which gives the indication that the QCD is an extended version of
electroweak theory \cite{hep-ph/0410207}. The quarks and leptons family
structure is presented in Ref. \cite{hep-ph/0501103} along with the nature
of all the four forces, i.e. Casimir force, gravitational force, electroweak
force and strong force. This gives the unified nature of all the four
forces. Where did we make mistake ?, is elaborated in Refs. \cite
{hep-ph/0502055,hep-ph/0502117}. Mass relations among quark families and
among lepton families are discussed in Ref. \cite{hep-ph/0503025}.

In the present study we obtain the various mass relations among members of
quark and lepton families. The mass relations for the charm, beauty, and top
quark family members are 
\begin{eqnarray*}
\left. m_{c^0}:m_{c^r}:m_{c^g}:m_{c^b}:m_{c^z}:m_{c^{\bar{b}}}:m_{c^{\bar{g}%
}}:m_{c^{\bar{r}}}\right. &=&\left. \frac{\sqrt{3}}4:\!\!\!\frac{\sqrt{19}}4:%
\frac{\sqrt{31}}4:\frac{\sqrt{7}}4:\frac{\sqrt{35}}4:\frac{\sqrt{19}}4:\frac{%
\sqrt{7}}4:\frac{\sqrt{31}}4\right. , \\
\left. m_{b^0}:m_{b^r}:m_{b^g}:m_{b^b}:m_{b^z}:m_{b^{\bar{b}}}:m_{b^{\bar{g}%
}}:m_{b^{\bar{r}}}\right. &=&\left. \frac 12:\frac 32:\frac{\sqrt{3}}2:\frac{%
\sqrt{3}}2:\frac 32:\frac 12:\frac{\sqrt{7}}2:\frac{\sqrt{7}}2\right. , \\
\left. m_{t^0}:m_{t^r}:m_{t^g}:m_{t^b}:m_{t^z}:m_{t^{\bar{b}}}:m_{t^{\bar{g}%
}}:m_{t^{\bar{r}}}\right. &=&\left. \frac 1{\sqrt{2}}:\sqrt{\frac 32}:\sqrt{%
\frac 32}:\sqrt{\frac 32}:\frac 3{\sqrt{2}}:\sqrt{\frac 32}:\sqrt{\frac 32}:%
\sqrt{\frac 32}\right. .
\end{eqnarray*}
The mass relations for the electron, muon and tau lepton family members are 
\begin{eqnarray*}
\left. m_{e^0}:m_{e^r}:m_{e^g}:m_{e^b}:m_{e^z}:m_{e^{\bar{b}}}:m_{e^{\bar{g}%
}}:m_{e^{\bar{r}}}\right. &=&\!\!\!\left. \frac{\sqrt{3}}4:\frac{\sqrt{19}}4:%
\frac{\sqrt{7}}4:\frac{\sqrt{31}}4:\frac{\sqrt{35}}4:\frac{\sqrt{7}}4:\frac{%
\sqrt{31}}4:\frac{\sqrt{19}}4\right. , \\
\left. m_{\mu ^0}:m_{\mu ^r}:m_{\mu ^g}:m_{\mu ^b}:m_{\mu ^z}:m_{\mu ^{\bar{b%
}}}:m_{\mu ^{\bar{g}}}:m_{\mu ^{\bar{r}}}\right. \!\! &=&\left. \frac
12:\frac 12:\frac{\sqrt{7}}2:\frac{\sqrt{7}}2:\frac 32:\frac{\sqrt{3}}2:%
\frac{\sqrt{3}}2:\frac 32\right. , \\
\left. m_{\tau ^0}:m_{\tau ^r}:m_{\tau ^g}:m_{\tau ^b}:m_{\tau ^z}:m_{\tau ^{%
\bar{b}}}:m_{\tau ^{\bar{g}}}:m_{\tau ^{\bar{r}}}\right. \!\!\! &=&\left.
\frac 1{\sqrt{2}}:\sqrt{\frac 32}:\sqrt{\frac 32}:\sqrt{\frac 32}:\frac 1{%
\sqrt{2}}:\sqrt{\frac 32}:\sqrt{\frac 32}:\sqrt{\frac 32}\right. .
\end{eqnarray*}

Many of friends and colleagues asked a question: What will you say about the
experimental facts which are explained with the help of different theories,
like QCD etc ? I can only resemble an accurate theory like an accurate
balance. If a balance is not accurate, how can we weigh with its help
accurately ? The similar will be the case with an approximated theory, you
can predict few of the experimental results with the help of one theory and
for the rest of facts you need other theories and so on. There will be no
end of new theories for every new phenomena. But an accurate theory will
solve all the problems and predict the data accurately. Massless QCD is an
approximated theory which based on the assumption $3^2-1=8$ and its draw
backs we have already discussed in Ref. \cite{hep-ph/0502117}. The
experimental results \cite
{PRL84-5283,PRL89-231801,hep-ex/0308021,hep-ex/0408138} will never be
explained correctly with the help of such approximated theories. Then, to
approach experimental results we need higher order corrections \cite
{PLB470-223,NPB643-431,hep-ph/0211381,hep-ph/0312090,PRD48-5196,hep-ph/0401038,PRD54-3350,hep-ph/0105302,PRD69-114007,hep-ph/0503039,hep-ph/0503095}%
. Even then we are unable to predict the exact experimental results \cite
{hep-ph/0409133}. A fundamental theory that we believe offers an extremely
complete and accurate set of equations governing the structure and behavior
of matter in ordinary circumstances.

\end{document}